%% file: nanolett.tex
\documentclass[journal=nalefd,manuscript=letter]{achemso}
\usepackage[version=3]{mhchem} 
\author{Masazumi Fujiwara}
\affiliation{Research Institute for Electronic Science, Hokkaido University, Sapporo, Hokkaido 001-0021, Japan}
\alsoaffiliation{The Institute of Scientific and Industrial Research, Osaka University, Ibaraki, Osaka 567-0047, Japan}
\author{Kiyota Toubaru}
\affiliation{Research Institute for Electronic Science, Hokkaido University, Sapporo, Hokkaido 001-0021, Japan}
\alsoaffiliation{The Institute of Scientific and Industrial Research, Osaka University, Ibaraki, Osaka 567-0047, Japan}
\author{Tetsuya Noda}
\affiliation{Research Institute for Electronic Science, Hokkaido University, Sapporo, Hokkaido 001-0021, Japan}
\alsoaffiliation{The Institute of Scientific and Industrial Research, Osaka University, Ibaraki, Osaka 567-0047, Japan}
\author{Hong-Quan Zhao}
\affiliation{Research Institute for Electronic Science, Hokkaido University, Sapporo, Hokkaido 001-0021, Japan}
\alsoaffiliation{The Institute of Scientific and Industrial Research, Osaka University, Ibaraki, Osaka 567-0047, Japan}
\author{Shigeki Takeuchi}
\email{takeuchi@es.hokudai.ac.jp}
\affiliation{Research Institute for Electronic Science, Hokkaido University, Sapporo, Hokkaido 001-0021, Japan}
\alsoaffiliation{The Institute of Scientific and Industrial Research, Osaka University, Ibaraki, Osaka 567-0047, Japan}

\title[\texttt{achemso} demonstration]
{Highly efficient coupling of photons from nanoemitters into single-mode optical fibers}

\keywords{tapered fiber, nanocrystal, single-photon source, coupling efficiency, high numerical aperture, photon collection optic}

\begin{document}
\begin{abstract}
Highly efficient coupling of photons from nanoemitters into single-mode optical fibers is demonstrated using tapered fibers. 
$7.4 \pm 1.2$ \% of the total emitted photons from single CdSe/ZnS nanocrystals were coupled into a 300-nm-diameter tapered fiber. 
The dependence of the coupling efficiency on the taper diameter was investigated and
the coupling efficiency was found to increase exponentially with decreasing diameter.
This method is very promising for nanoparticle sensing and single-photon sources.
\end{abstract}

Collection of fluorescence photons from single nanoemitters is of fundamental importance 
in fields such as quantum information and biological sensing.
For examples, single nanoemitters such as quantum dots (QDs) and color defect centers in nanodiamonds can be employed as single-photon sources, which are crucial devices for realizing quantum cryptography in future secure communication networks
\cite{SPS:Moerner2000, santori2002indistinguishable, pelton2002PRL, sanaka2010PRL, 
Qdot_interfere_PRL2010, diamond_SPS1, diamond_SPS2}. 
Fiber-based sensing of fluorescent nanoparticles with a ultrahigh sensitivity has also been intensively studied \cite{fiber-sensing-nanodevice1, fiber-sensing-nanodevice2}. 
All these applications require the ability to collect as many photons as possible and to efficiently couple these photons into single-mode optical fibers.

To enhance the fluorescence collection efficiency of photons from these nanoemitters, 
various photon collection optics have been investigated, 
including solid immersion lenses \cite{tim_SIL, jeremy_SIL}, photonic crystal fibers\cite{tim_pcfiber}, 
and tapered fibers\cite{taper_SPS, nayak2007optical, davanco2009efficient}. 
Tapered fibers are particularly promising in view of their high collection efficiencies and their ability to directly couple fluorescence photons into a single-mode fiber.
A theoretical study has predicted that it should be possible to couple 28 \% of the total emission from gas atoms around the taper region 
into single-mode fiber outputs \cite{kien_PRA}. 
Preliminary experimental results for coupling between tapered fibers and 
solid-state nanoemitters have been reported \cite{gregor2009soft, stiebeiner2009ultra, srinivasan2007single}.
However, efficient coupling of fluorescence from single solid-state nanoemitters into tapered fibers 
has not been reported. This is mainly due to the following two difficulties: 
(1) Efficient coupling of nanoemitters into tapered fibers requires ultrasmall taper diameters 
of the order of 300 nm (approximately half the emission wavelength) as in Ref.\citenum{kien_PRA}; however, it is currently challenging to fabricate such ultrathin tapered fibers with low transmission loss \cite{hartung2010limits}. 
(2) Ultrathin tapered fibers suffer from rapid transmission degradation \cite{hokori_opex_2011}.

In this study, we demonstrate highly efficient coupling of fluorescence from single QDs 
into single-mode fibers by using ultrathin tapered fibers.
We succeeded in producing tapered fibers with a diameter of 300 nm and a transmittance of 90 \%.
We preserved their transmittance by conducting all experiments in a dust-free environment.
We were able to couple $7.4 \pm 1.2$ \% of the total emitted photons from single CdSe/ZnS nanocrystals into tapered fibers.
This efficient photon collection technique is highly promising 
for nanoparticle sensing and single-photon sources.
\begin{figure}[t!]
	\includegraphics{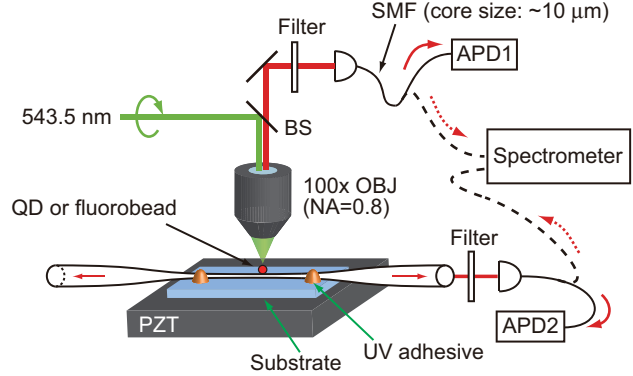}
	\caption{(Color online) A schematic diagram of the experimental setup. 
	OBJ: objective; BS: dichroic beam splitter; SMF: single-mode fiber; APD: avalanche photodiode.
	The tapered fiber was fixed to a glass substrate using UV adhesives and the substrate was mounted on the PZT.
	To measure fluorescence spectra, the fiber connections to the APDs were switched to spectrometer.}
\label{fig1} 
\end{figure}

Tapered fibers were fabricated from standard single-mode optical fibers (Thorlabs, 630HP) 
by a procedure described elsewhere \cite{chiba_apl, konishi_apl, tanaka_opex, hokori_opex_2011}. 
By making the taper transition region 5 mm longer than in previous studies \cite{hokori_opex_2011}, 
we achieved a taper diameter of 300 nm with a transmittance of greater than 0.9. 
The transmittances of all tapered fibers used (diameters: 0.3--1.0 $\mu$m) were measured 
during fabrication and they were confirmed to be greater than 0.9.
Their diameters were measured by scanning electron microscopy prior to the optical experiments.
The tapered fibers were stored in a dust-free environment (class 10 cleanroom) until use
to minimize degradation of their transmittance \cite{hokori_opex_2011}.
All experiments were performed in a class 100 cleanroom 
for which transmission degradation was negligible during the optical experiment.
We used CdSe/ZnS QDs (Evident; crystal size: 9.6 nm; maximum emission wavelength: 620 nm; 
fluorescence quantum efficiency: 0.5) and fluorobeads (Molecular Probes, F8887; size: 200 nm) 
as solid-state nanoemitters.
The QDs were dissolved in toluene solution, while the fluorobeads were dispersed in 2-ethoxyethanol.
The tapered fibers were dipped in these solutions to deposit these nanoemitters directly on their tapered surfaces 
and they were then mounted on a piezoelectric transducer (PZT).
This dip coating did not reduce the transmittances of the tapered fibers by more than 10 \%. 

Figure \plainref{fig1} schematically depicts the experimental setup. 
A green He--Ne laser beam ($\lambda$ = 543.5 nm) was used as the excitation light. 
It was circularly polarized to eliminate the polarization dependence of the fluorescence intensity 
of single QDs. It was focused on the tapered fiber by a 100x objective whose numerical aperture (NA) was 0.8.
The laser spot size was confirmed to be diffraction limited.
The fluorescence was collected by the same objective and it was filtered from the excitation laser light using a dichroic beam splitter and additional filters. 
Since the optimal pinhole size for confocal detection was 10--20 $\mu$m, 
we used a single-mode optical fiber whose core size was approximately 10 $\mu$m (Thorlabs, 1550HP) 
instead of a pinhole. 
The fluorescence was then detected by an avalanche photodiode (denoted APD1).
The fluorescence channeled into the tapered fiber was detected by another APD (APD2) 
after filtering the excitation light.
In this experiment, the fluorescence was directed to the both ends of tapered fibers with a ratio of 50:50; 
only one side of the fluorescence was detected using this experimental setup.
We experimentally confirmed that the same amount of fluorescence photons 
were detected at both ends of the tapered fiber.
The two APDs (APD1 and APD2) together with a time-correlated single photon counting module (Becker \& Hickl, SPC-130) 
were used for second-order photon correlation measurements in order to discriminate single emitters
by observing the antibunching correlation.
The fiber connections to the APDs were switched to the spectrometer to measure the fluorescence spectra.

\begin{figure}[t!]
	\includegraphics{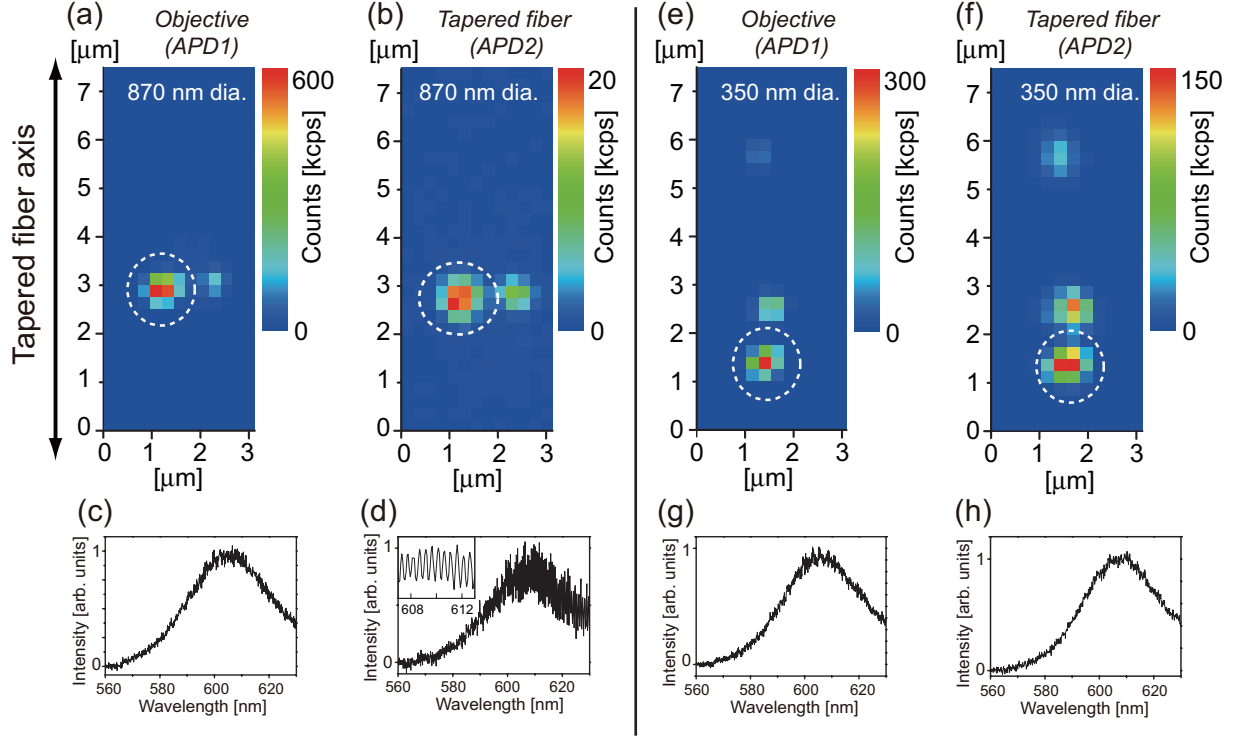}
	\caption{(Color online) Scanning images of fluorobeads deposited on the 870-nm-diameter tapered fiber, 
	in which the fluorescence was detected through (a) the objective and (b) the tapered fiber, respectively.
	In each figure, there were two beads observed on the fiber, 
	and the bead indicated by white circle was used to obtain the fluorescence spectrum.
	Fluorescence spectra of the bead on the 870-nm fiber measured through (c) the objective and (d) the tapered fiber.
	Scanning images of beads on 350-nm fiber obtained through (e) the objective and (f) the tapered fiber, respectively.
	In each figure, three beads were observed along the fiber axis, 
	and the bead in the white circle was used for the fluorescence spectral measurements.
	Fluorescence spectra of the bead on 350-nm fiber measured through (g) the objective and (h) the tapered fiber.
	The tapered fiber axis goes from top to bottom in (a), (b), (e), and (f).
	}
	\label{fig2}
\end{figure}

We first tested coupling of the fluorobeads with a relatively thick tapered fiber whose 
diameter was 870 nm.
Figures \plainref{fig2}(a) and (b) show scanning images of fluorobeads on this fiber
obtained through the objective (APD1) and the tapered fiber (APD2), respectively.
Both methods detected the same fluorobeads on the fiber.
Focusing on the bead indicated by the white circles, 
the photon counts detected through the tapered fiber was 20 kcps 
(\textit{i.e.}, 40 kcps for fiber-coupled photons since we only measured the photons from one of the two fiber ends).
This value of 40 kcps is 15 times smaller than that (600 kcps) obtained through the objective. 

Figures \plainref{fig2}(c) and (d) show fluorescence spectra of this bead 
measured through the objective and the tapered fiber, respectively. 
These fluorescence spectra differ significantly from each other: 
an interference structure appeared in the spectrum measured through the tapered fiber [Fig. \plainref{fig2}(d)].
This interference structure originates from multimode interference in the tapered fiber.
It is well known that transmittance spectra of thick tapered fibers exhibit this kind of structure 
due to multimode interference \cite{mmi1}. 
We performed numerical calculations using a waveguide solver (Photon Design, Fimmwave Software) and 
they indeed predicted three propagation modes in the 870-nm tapered fiber, 
which is consistent with the multimode nature of the fluorescence spectrum.

In the same way, we obtained scanning images of beads on a 350-nm fiber. 
Figures \plainref{fig2}(e) and (f) show images obtained through the objective and the tapered fiber, respectively.
The photon count of the bead measured through the tapered fiber increased to 150 kcps 
(\textit{i.e.}, 300 kcps for the total amount of fluorescence photons coupled into the tapered fiber).
This value of 300 kcps is comparable to the photon count obtained through the objective, 
namely by a conventional confocal microscope setup [Fig. \plainref{fig2}(e)].
Figures \plainref{fig2}(g) and (h) show fluorescence spectra of the same bead measured through 
the objective and the tapered fiber, respectively.
The interference structure in Fig. \plainref{fig2}(d) is not present in Fig. \plainref{fig2}(h), 
indicating that only a single propagating mode was coupled with the fluorobeads.

\begin{figure}[t!]
	\includegraphics{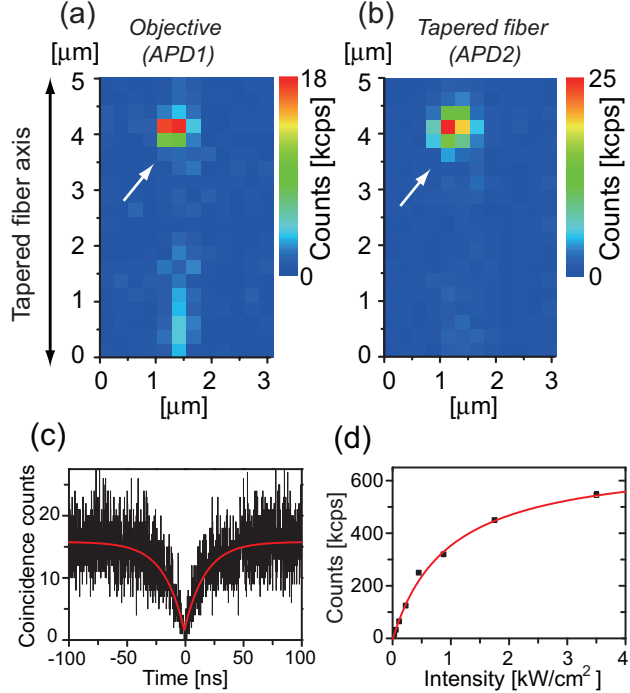}
	\caption{(Color online) Scanning images of QDs on tapered fiber measured  through 
	(a) objective and (b) tapered fiber.
	The excitation intensity was 28 W/cm$^2$. 
	A single QD (indicated by white arrows) is visible in both images.
	The low-intensity blurred spots appeared in the lower part of (a) are other QDs, 
	which are not clear in (b) simply because their relative intensity to the photon count of the brightest spot is smaller in (b).
	(c) Histogram of the second-order photon correlation of this QD.
	(d) Excitation intensity dependence plot of the photon counts of this QD. 
	}
	\label{fig3}
\end{figure}

Next, we demonstrate highly efficient coupling of the fluorescence from single QDs to ultrathin (300-nm) tapered fibers.
Figures \plainref{fig3}(a) and (b) show scanning images of a single QD on a 300-nm-diameter tapered fiber, 
which were measured through the objective and the tapered fiber, respectively.
We observed single-step fluorescence blinking at this spot (data not shown).
The fluorescence detected [25 kcps in Fig. \plainref{fig3}(b)] through the tapered fiber was larger than 
that [18 kcps in Fig. \plainref{fig3}(a)] detected through the objective.
The total photons that could be detected from both ends of the tapered fiber was 50 kcps, 
which is almost three times greater than the photon count detected through the objective.
Figure \plainref{fig3}(c) shows a second-order photon correlation histogram of this single QD, 
where the coincidence events between APD1 (through the objective) and APD2 (through the fiber) 
were recorded in terms of time difference.
It clearly shows antibunching at time 0 with a value of $g^{(2)}(0) = 0.096$, 
demonstrating that it is a single emitter \cite{Moerner:CPL2000}.
The excited-state lifetime of this QD was determined to be 35.4 ns from 
this second-order photon correlation data. 
Figure \plainref{fig3}(d) shows an excitation-intensity dependence plot of the fluorescence detected by APD2. 
It shows that the fluorescence saturates as the excitation intensity increased.
We measured the second-order photon correlation and fluorescence saturation 
for three other single QDs on the same 300-nm tapered fiber 
and found that the mean lifetime ($\tau _{\rm qd}$) and mean saturated photon counts ($C_{\infty}$) 
were $\tau _{\rm qd} = 29.6\pm2.1$ns 
(in agreement with previously reported values\cite{Moerner:CPL2000, de2003single}) 
and $C_{\infty} = 592\pm90$ kcps, respectively. 

Given the experimentally obtained values for $\tau _{\rm qd}$ and $C_{\infty}$, 
we can calculate the coupling efficiency of the QD fluorescence to the tapered fiber ($\eta $) 
using the following equation 
\begin{equation}
\eta = \frac{2 C_{\infty} \tau _{\rm qd}}{\chi  _D T}, 
\label{eq1}
\end{equation}
where $\chi_D$ is the APD quantum efficiency 
($\chi_D = 0.68$ at $\lambda = 600$ nm according to manufacturer's specification sheet) 
and $T$ is the measured transmission loss from the output of the tapered fiber to APD2 ($T = 0.7$). 
The factor of $2$ in \ref{eq1} originates from the experimental conditions and we confirmed experimentally 
that the fluorescence was equally coupled 
to the two propagation directions in the tapered fiber.
Hence, we obtained $\eta = 7.4 \pm 1.2$ \% for the total coupling efficiency of the QD fluorescence into the tapered fiber.

\begin{figure}[t!]
	\includegraphics{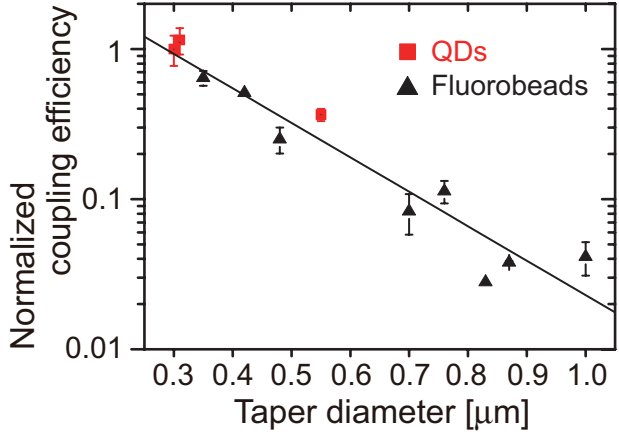}
	\caption{(Color online) Dependence of the fluorescence coupling efficiency on the taper diameter. 
	The coupling efficiency is normalized to that at a taper diameter of 300 nm. 
	Black triangles and red rectangles denote fluorobeads and QDs, respectively.
	The solid line is a single exponential fit to the data.
	}
	\label{fig4}
\end{figure}

We also investigated the dependence of the fluorescence coupling efficiency on the taper diameter.
We measured the fluorescence photon count of single QDs and fluorobeads through the objective 
and tapered fiber for various taper diameters. 
Figure \plainref{fig4} shows the dependence of the coupling efficiency on the taper diameter.
The coupling efficiency is normalized to that at a taper diameter of 300 nm. 
As the taper diameter decreases, the coupling efficiency increases exponentially. 
This diameter dependence is consistent with the fact that the intensity of the evanescent field, 
which is the one coupled with the emitters, decreases exponentially 
outside the tapered fiber\cite{kien_PRA, nayak2007optical}.

The observed saturated photon counts for single QDs 
are larger than that reported in previous reports using a standard confocal microscope setup 
\cite{Moerner:CPL2000, brokmann2004highly}. 
For example, Lounis \textit{et al.} reported $\sim$ 600 kcps (deduced from figure) 
for the saturation photon count of a single CdSe/ZnS QD 
by using a 1.4-NA oil immersion objective\cite{Moerner:CPL2000}.
When we take into account the fact that the number of photons coupled into the tapered fiber 
is twice the observed photon count, 
the saturated photon count observed in the tapered fiber may be $\sim$1200 kcps 
(the observed one is $C_{\infty} = 592\pm90$ kcps), 
which is approximately two times larger than that obtained using the oil-immersion objective.
Note that it is possible to extract all the photons from the one end of the fiber 
by connecting the one end with a pigtailed mirror or a fiber-Bragg grating.
Our results demonstrate a single QD-tapered fiber integrated single-photon source 
with a potential maximum photon detection rate of 1.2 MHz.

The most prominent characteristic of this single-photon source is that 
trains of single photons are output from conventional single-mode optical fibers. 
Coupling single-photon sources to single-mode optical fibers is critical for quantum optics experiments, 
such as the observation of two-photon interference, which is essential for quantum cryptography 
\cite{santori2002indistinguishable, Qdot_interfere_PRL2010}. 
One possible application of the present tapered fiber single-photon source is quantum key distribution. 
Such a single photon source will be an alternative candidate to the reported heralding single photon sources 
with the efficiencies of 20 to 30 \% \cite{soujaeff-JOptMod, soujaeff2007quantum}. 
Another future target will be optical quantum computation, where the loss threshold of 66 \% 
has been suggested for a certain condition \cite{VarnavaPRL}.

Efficient photon collection of single nanoemitters itself is possible also by using 
high-NA objectives such as solid immersion lenses (SILs).
18.3 \% of total emission from fluorescent diamond nanocrystals, which were placed on the SIL surface, 
has been reported \cite{tim_SIL}.
However, subsequent optical components such as objectives and single-mode fibers reduce 
the actual collection efficiency down to 1--2 \% of the total emission from single nanoemitters \cite{toishi2009high}.
Thus, the present coupling efficiency of 7.4 \% for coupling of fluorescence into single-mode fiber outputs is promising 
for future quantum information devices.
Note that alternative approaches to enhance the coupling efficiency 
by embedding single nanoemitters into nano-optical structures 
have been recently reported \cite{claudon2010highly, stephan_nphoton}.

Tapered fibers also have the following two promising characteristics:
(1) They can be used at cryogenic temperatures\cite{takashima2010fiber}, which are necessary for suppressing the 
thermal contribution to the fluorescence linewidth of solid-state nanoemitters
\cite{santori2002indistinguishable, Qdot_interfere_PRL2010}. 
(2) They can be applied to nanostructural single-photon sources 
such as micropost microcavities \cite{pelton2002PRL}, diamond nanowires \cite{diamond_SPS2}, 
and diamond nanopillars \cite{hwang2009composite}, 
which cannot be accessed by oil- or solid-immersion high-NA objectives.
For these reasons, tapered fibers are highly promising photon collection optics 
that permit highly efficient direct coupling of nanoemitter fluorescence into single-mode optical fibers. 

In conclusion, highly efficient fluorescence coupling from single QDs into single-mode optical 
fibers has been demonstrated using ultrathin tapered fibers.
$7.4 \pm 1.2$ \% of the total emitted photons from single CdSe/ZnS nanocrystals were coupled into 300-nm-diameter tapered fibers.
The dependence of the coupling efficiency on the taper diameter was investigated.
The coupling efficiency was found to increase exponentially with decreasing diameter.
This highly efficient photon collection technique has great potential for 
nanoparticle sensing and single-photon sources.

\acknowledgement
We gratefully acknowledge financial support from MIC-SCOPE, JST-CREST, 
MEXT-KAKENHI Quantum Cybernetics (No. 21102007), 
JSPS-KAKENHI (Nos. 20244062, 21840003, 23244079, and 23740228), 
JSPS-FIRST, Project for Developing Innovation Systems of MEXT, 
G-COE program, and Research Foundation for Opto-Science and Technology.

\input{nanolett.bbl}

\end{document}

%% file: nanolett.bbl
\providecommand*\mcitethebibliography{\thebibliography}
\csname @ifundefined\endcsname{endmcitethebibliography}
  {\let\endmcitethebibliography\endthebibliography}{}